\documentstyle[epsfig]{article}
\title{Quantum FRW cosmological solutions in the presence of Chaplygin gas and perfect fluid}
\author{P.~Pedram\thanks{Email: pouria.pedram@gmail.com}, S.~Jalalzadeh\thanks{Email: s-jalalzadeh@sbu.ac.ir},
\\ {\small Department of Physics, Shahid Beheshti University,
Evin, Tehran 19839, Iran}}
\begin{document}
\maketitle \baselineskip 24pt
\begin{abstract}
We present a Friedmann-Robertson-Walker quantum cosmological model
in the presence of Chaplygin gas and perfect fluid for early and
late time epoches. In this work, we consider perfect fluid as an
effective potential and apply Schutz's variational formalism to the
Chaplygin gas which recovers the notion of time. These give rise to
Schr\"odinger-Wheeler-DeWitt equation for the scale factor. We use
the eigenfunctions in order to construct wave packets and study the
time dependent behavior of the expectation value of the scale factor
using the many-worlds interpretation of quantum mechanics. We show
that contrary to the classical case, the expectation value of the
scale factor avoids singularity at quantum level. Moreover, this
model predicts that the expansion of Universe is accelerating for the
late times.
\end{abstract}

\textit{Pacs}:{ 98.80.Qc, 04.40.Nr, 04.60.Ds}

\section{Introduction}\label{sec1}
Supernova Ia (SNIa) observations show that the expansion of the
Universe is accelerating \cite{Riess:1998cb}, contrary to
Friedmann-Robertson-Walker (FRW) cosmological models, with
non-relativistic matter and radiation. Also cosmic microwave
background radiation (CMBR) data \cite{Spergel:2003cb,2a} is
suggesting that the expansion of our Universe seems to be in an
accelerated state which is referred to ``dark energy'' effect
\cite{3a}. Cosmological constant, $\Lambda$, as the vacuum energy
can be responsible for this evolution by providing a negative
pressure \cite{3b,3c}. Unfortunately, the observed value of
$\Lambda$ is $120$ orders of magnitude smaller than the one computed
from field theory models \cite{3b,3c}. Quintessence is an
alternative to consider a dynamical vacuum energy
\cite{Wetterich:fm}, involving one or two scalar fields, some with
potentials justified from supergravity theories \cite{Brax:1999yv}.
However, the fine-tuning problem of these models which arises from
cosmic coincidence issue has no satisfactory solution.

The Chaplygin gas model is an interesting proposal
\cite{Kamenshchik}, describing a transition from a Universe filled
with dust-like matter to an accelerating expanding stage. This model
was later generalized in Ref.~\cite{Kamenshchik,A}. The generalized
Chaplygin gas model is described by a perfect fluid obeying an
exotic equation of state \cite{Kamenshchik,A}
\begin{equation}
p=-\frac{A}{\rho ^{\alpha }},  \label{cgi1}
\end{equation}
where $A$ is a positive constant and $0<\alpha \leq 1$. The standard
Chaplygin gas \cite{Kamenshchik} corresponds to $\alpha =1$. Some
publications
\cite{Bento,3rp,2rp,Fabris,C,Ogawa,NewBD,18a,20a,21a,23a,27a,Rev1,Rev2,DySy,Jackiw,setare1,setare2,setare3}
and reviews \cite{Rev1,Rev2} which studied the Chaplygin gas
cosmological models have already appeared in the literature. The
Chaplygin gas can be obtained from the string Nambu-Goto action in
the light cone coordinate \cite{Jackiw}. Since the application of
string theory in principle is in very high energy when the quantum
effects is important in early Universe, a quantum cosmological study
of the Chaplygin gas is also well founded.

Recently, Quantum mechanical description of a FRW model with a
generalized Chaplygin gas has been discussed in Ref.~\cite{Buahmadi}
in order to retrieve explicit mathematical expressions for the
different quantum mechanical states and determine the transition
probabilities towards an accelerated stage. Moreover, quantization
of FRW model in the presence of Chaplygin gas has been discussed in
Ref.~\cite{chap}. There, we have considered matter as the Chaplygin
gas and discussed the early time behavior of expectation value of
the scale factor through the application of Schutz's formalism. In
this paper, aside from the Chaplygin gas which is coupled to gravity
and has an advantage of furnishing a variable connected to matter
which can be identified with time, we also include the perfect fluid
in this scenario and investigate the analytical solutions in both
early and late time Universes. Schutz's formalism \cite{11,12} gives
dynamics to the matter degrees of freedom in interaction with the
gravitational field. Using proper canonical transformations, at
least one conjugate momentum operator associated with matter appears
linearly in the action integral. Therefore, a Schr\"odinger-like
equation can be obtained with the matter variable playing the role
of time. The application of Schutz's formalism in Stephani and FRW
perfect fluid cosmological models has been discussed in
Refs.~\cite{PLB,pedramCQG2,FRW}. Note that our approach in principle
is different from Monerat \textit{et. al} \cite{monerat2}, where
they correspond the dynamical variable to the perfect fluid instead
of Chaplygin gas and resort to the numerical methods to obtain the
time evolution of an initial wave packet. Note that there is
considerable evidence that the early Universe is dominated by
radiation. Therefore, a natural setting for quantum cosmology is the
one where radiation has the predominant role \cite{ref1}. On the
other hand the Chaplygin gas is dominated by the non-relativistic
matter at early times (See the following section). This seems to be
in contradiction with our knowledge of baby Universe. According to
\cite{ref2}, inflation can be accommodated within the generalized
Chaplygin gas scenario. Hence, the way adopted to avoid this
inconsistency is that the radiation dominated phase is followed by
Chaplygin dominated period so that we have the so-called Chaplygin
inflation \cite{ref2}. Also, it would be more suitable to consider
the field theory representation of the Chaplygin gas \cite{ref3} to
describe quantum cosmology. In this way, the Chaplygin gas can be
viewed as a modification of gravity as was first pointed out in
\cite{ref3}. Also, the authors of \cite{ref4} have recently shown that the
Chaplygin gas model has a geometrical explanation within the context
of brane world theory for any $\alpha$. Consequently in these models
the equation
\begin{eqnarray}\label{ref}
\rho
=\left[A+\frac{B}{a(t)^{3(\alpha+1)}}\right]^{\frac{1}{1+\alpha}},
\end{eqnarray}
is a consequence of stress-energy conservation for a scalar field on
the brane \cite{ref3}, or conservation of induced dark matter on the
brane \cite{ref4,ref5}. Here, $a(t)$ is the scale-factor of the
universe and $B$ is a positive integration constant. Therefore, it
is evocative to view the contribution of Chaplygin gas to the
stress-energy tensor as a brane induced modification of gravity. In
this article, we used the fluid description for the Chaplygin gas
and for Lagrangian formalism, the corresponding pressure.
Consequently, if we rely on the model described in \cite{ref4}, we
will have covariance in our model.

The paper is organized as follows. In Sec.~\ref{sec2}, the quantum
cosmological model with a Chaplygin gas, as a portion of the matter
content is constructed in Schutz's formalism \cite{11,12} for early
and late time Universes. Then the Schr\"odinger-Wheeler-DeWitt (SWD)
equation in minisuperspace is obtained to quantize the model under
the action of a perfect fluid effective potential. The wave function
depends on the scale factor $a$ and on the canonical variable
associated to the Chaplygin gas, which in the Schutz's variational
formalism plays the role of time $T$. We separate the wave function
into two parts, one depends only on the scale factor and the other
depends on the time. The time dependent part of the solution is
$e^{iEt}$, where $E$ is the energy. In Sec.~\ref{sec3}, we construct
wave packets from the eigenfunctions and compute the time-dependent
expectation values of the scale factor to investigate the existence
of singularities at quantum level. Moreover, we present some
analytical solutions in both early and late time epoches. In
Sec.~\ref{sec4}, we present our conclusions.

\section{The Model}\label{sec2}
The action for gravity plus Chaplygin gas in Schutz's formalism is
written as
\begin{eqnarray}
 S= \int_Md^4x\sqrt{-g}\, R + 2\int_{\partial
M}d^3x\sqrt{h}\, h_{ab}\, K^{ab}+ \int_Md^4x\sqrt{-g}\,\, p_f +
\int_Md^4x\sqrt{-g}\,\, p_c,\label{action}
\end{eqnarray}
here, $K^{ab}$ is the extrinsic curvature and $h_{ab}$ is the
induced metric over the three-dimensional spatial hypersurface,
which is the boundary $\partial M$ of the four dimensional manifold
$M$. We choose units such that the factor $16\pi G$ becomes equal to
one. $p_f$ and $p_c$ denote Chaplygin gas and perfect fluid
pressure, respectively. Note that, according to \cite{11} the above
action is equivalent to the usual Hawking-Ellis formalism for
perfect fluid description \cite{ref7}. Perfect fluid satisfies the
barotropic equation of state
\begin{eqnarray}
 p_f=w\rho_f,\quad\quad w\leq1.
\end{eqnarray}
The first two terms were first obtained in \cite{7} and the last
term of (\ref{action}) represents the matter contribution to the
total action. In Schutz's formalism \cite{11,12} the fluid's
four-velocity can be expressed in terms of five potentials $\Phi$,
$\zeta$, $\beta$, $\theta$ and $S$
\begin{equation}
u_\nu = \frac{1}{\mu}(\Phi_{,\nu} + \zeta\beta_{,\nu} + \theta S_{,\nu})
\end{equation}
where $\mu$ is the specific enthalpy. $S$ is the specific entropy,
and the potentials $\zeta$ and $\beta$ are connected with rotation
which are absent of models in the Friedmann-Robertson-Walker (FRW)
type. The variables $\Phi$ and $\theta$ have no clear physical
meaning. The four-velocity also satisfies the normalization
condition
\begin{equation}
u^\nu u_\nu = -1.
\end{equation}
The FRW metric
\begin{equation}
ds^2 = - N^2(t)dt^2 + a^2(t)g_{ij}dx^idx^j,
\end{equation}
can be inserted in the action (\ref{action}), where $N(t)$ is the
lapse function and $g_{ij}$ is the metric on the constant-curvature
spatial section. Following the thermodynamic description of
Ref.~\cite{14}, the basic thermodynamic relations take the form
\begin{eqnarray}
% \nonumber to remove numbering (before each equation)
  \rho_c &=& \rho_0[1+\Pi], \quad h=1+\Pi+p_c/\rho_0, \\ \nonumber
  \tau dS &=&
  d\Pi+p_c\,d(1/\rho_0),\\
  &=&\frac{(1+\Pi)^{-\alpha}}{1+\alpha}d\left[(1+\Pi)^{1+\alpha}+\frac{A}{\rho_0^{1+\alpha}
  }\right].
\end{eqnarray}
It then follows that to within a factor
\begin{eqnarray}
% \nonumber to remove numbering (before each equation)
  \tau &=& \frac{(1+\Pi)^{-\alpha}}{1+\alpha}, \\
  S &=& (1+\Pi)^{1+\alpha}+\frac{A}{\rho_0^{1+\alpha}}.
\end{eqnarray}
Therefore, the equation of state takes the form
\begin{equation}
   p_c=-A\left[\frac{1}{A}\left(1-\frac{\,\,h^{\frac{1+\alpha}{\alpha}}}{S^{1/\alpha}}\right)\right]^{\frac{1+\alpha}{\alpha}}.
\end{equation}
The particle number density and energy density are, respectively,
\begin{eqnarray}
% \nonumber to remove numbering (before each equation)
  \rho_c &=& \left[\frac{1}{A}\left(1-\frac{\,\,h^{\frac{1+\alpha}{\alpha}}}{S^{1/\alpha}} \right) \right]^{\frac{-1}{\,\,1+\alpha}}, \\
  \rho_0 &=& \frac{\rho+p}{h},
\end{eqnarray}
where $h=(\dot{\Phi}+\theta\dot{S})/N$. After dropping the surface terms, the final reduced action takes the form
\begin{eqnarray}
S = \int dt\biggr\{-6\frac{\dot a^2a}{N} + 6kNa -N a^3 \rho_f -N a^3
A\left[\frac{1}{A}\left(1-\frac{\,\,(\dot{\Phi}+\theta\dot{S})^{\frac{1+\alpha}{\alpha}}}{N^{\frac{1+\alpha}{\alpha}}
S^{1/\alpha}}\right)\right]^{\frac{1+\alpha}{\alpha}}\biggr\}.
\end{eqnarray}
The reduced action may be further simplified using canonical methods
\cite{14}, resulting in the super-Hamiltonian
\begin{equation}\label{superH}
{\cal H} = - \frac{p_a^2}{24a} -6ka +a^3 \rho_f +\left(S
p_{\Phi}^{1+\alpha}+A a^{3(1+\alpha)} \right)^{\frac{1}{1+\alpha}},
\end{equation}
where $p_a= -12{\dot aa}/{N}$ and $p_\Phi =\frac{\displaystyle
\partial{\cal L}}{\displaystyle \partial \dot{\Phi}}\,$. However, an analytical
quantum mechanical treatment of this FRW minisuperspace with the
above Hamiltonian does not seem feasible. Therefore, it requires
some approximation. We study the Chaplygin gas expression in early
and late times limits, namely for small scale factors $S
p_{\Phi}^{1+\alpha}\gg A a^{3(1+\alpha)}$ \cite{Buahmadi,chap} and
large scale factors $S p_{\Phi}^{1+\alpha}\ll A a^{3(1+\alpha)}$,
separately. So for early Universe, we can use the following
expansion
\begin{eqnarray}
\big(S p_{\Phi}^{1+\alpha}+ A a^{3(1+\alpha)}\big)
^{\frac1{1+\alpha}}\approx S^{\frac{1}{1+\alpha}} p_{\Phi}\bigg[
1+\frac1{1+\alpha}\frac{Aa^{3(\alpha+1)}}{S p_{\Phi}^{1+\alpha}}
+\frac12\frac1{1+\alpha}\left( \frac1{1+\alpha}-1\right)
\frac{A^{2}}{S^2
p_{\Phi}^{2(1+\alpha)}}a^{6(\alpha+1)}+\ldots\bigg].
\end{eqnarray}
Hence, up to the leading order, the super-Hamiltonian takes the form
\begin{equation}
{\cal H} = - \frac{p_a^2}{24a}-6ka  +a^3
\rho_f+S^{\frac{1}{1+\alpha}} p_{\Phi}.
\end{equation}
The following additional canonical transformations
\begin{eqnarray}
T =-(1+\alpha)p_\Phi^{-1}  S^{\frac{\alpha}{1+\alpha}}p_S, \quad
\quad p_T =S^{\frac{1}{1+\alpha}} p_\Phi,
\end{eqnarray}
and use of the explicit form of the energy density of the perfect
fluid $\rho_f=\frac{\displaystyle B}{\displaystyle a^{3(1+w)}}$,
simplify the super-Hamiltonian to
\begin{equation}
{\cal H} = - \frac{p_a^2}{24a} -6ka  +Ba^{-3w}+
p_T,\label{EqHamiltonian}
\end{equation}
where $B$ is a constant and the momentum $p_T$ is the only remaining
canonical variable associated with matter. It appears linearly in
the super-Hamiltonian. The parameter $k$ defines the curvature of
the spatial section, taking the values $0, 1, - 1$ for a flat,
positive-curvature or negative-curvature Universe, respectively.

The classical dynamics is governed by the Hamilton equations,
derived from Eq. (\ref{EqHamiltonian}) and Poisson brackets as
\begin{equation}
\left\{
\begin{array}{llllll}
\dot{a} =&\{a,N{\cal H}\}=-\frac{\displaystyle Np_{a}}{\displaystyle 12a} ,\\
 & \\
\dot{p_{a}} =&\{p_{a},N{\cal H}\}=- \frac{N}{24a^2}p_a^2+6Nk+3wNBa^{-3w-1}, \\
& \\
\dot{T} =&\{T,N{\cal H}\}=N\, ,\\
 & \\
\dot{p_{T}} =&\{p_{T},N{\cal H}\}=0\, .\\
& \\
\end{array}
\right. \label{4}
\end{equation}
We also have the constraint equation ${\cal H} = 0$. Choosing the
gauge $N=1$, we have the following solutions for the system
\begin{eqnarray}\label{class1}
T&=&t,\\\label{class2} p_T&=&\textrm{const.},\\\label{class3}
\ddot{a}&=&-\frac{\dot
a^2}{2a}-\frac{k}{2a}-\frac{1}{4}wBa^{-3w-2},\\\label{class4}
0&=&-6a\dot a^2-6k a +Ba^{-3w}+\,p_T.
\end{eqnarray}
The WD equation in minisuperspace can be obtained by imposing the
standard quantization conditions on the canonical momenta
($p_a=-i\frac{\displaystyle
\partial}{\displaystyle \partial a}$, $p_T=-i\frac{\displaystyle
\partial}{\displaystyle \partial T}$ ) and demanding that the
super-Hamiltonian operator annihilate the wave function ($\hbar =1$)
\begin{equation}
\label{sle} \frac{\partial^2\Psi}{\partial a^2} -
(144ka^2-24Ba^{1-3w})\Psi - i24a\frac{\partial\Psi}{\partial t} = 0.
\end{equation}
In this equation according to (\ref{class1}), $T=t$ corresponds to
the time coordinate. As discussed in \cite{nivaldo,15}, in order for
the Hamiltonian operator ${\hat H}$ to be self-adjoint  the inner
product of any two wave functions $\Phi$ and $\Psi$ must take the
form
\begin{equation}\label{inner}
(\Phi,\Psi) = \int_0^\infty a\,\Phi^*\Psi da,
\end{equation}
On the other hand, the wave functions should satisfy the following
boundary conditions
\begin{equation} \label{boundary} \Psi(0,t) = 0
\quad \mbox{or} \quad \frac{\partial\Psi (a,t)}{\partial
a}\bigg\vert_{a = 0} = 0.
\end{equation}
The SWD equation (\ref{sle}) can  be solved by separation of
variables as follows
\begin{equation}
\psi(a,t) = e^{iEt}\psi(a), \label{11}
\end{equation}
where the $a$ dependent part of the wave function $\psi(a)$
satisfies
\begin{equation}
\label{sle2} -\psi''(a) +(144 ka^2-24Ba^{1-3w})\psi(a)
=24Ea\,\psi(a),
\end{equation}
and the prime means derivative with respect to $a$.

Now, we consider late time Universe when $S p_{\Phi}^{1+\alpha}\ll A
a^{3(1+\alpha)}$. Using the expression
\begin{eqnarray}
\big(S p_{\Phi}^{1+\alpha}+ A a^{3(1+\alpha)}\big)
^{\frac1{1+\alpha}}\approx A^{\frac{1}{1+\alpha}} a^3\bigg[
1+\frac1{1+\alpha}\frac{S p_{\Phi}^{1+\alpha}}{ Aa^{3(\alpha+1)}}
+\frac12\frac1{1+\alpha}\left( \frac1{1+\alpha}-1\right) \frac{S^2
p_{\Phi}^{2(1+\alpha)}}{A^{2}a^{6(\alpha+1)}}+\ldots\bigg],
\end{eqnarray}
up to the first order, the super-Hamiltonian (\ref{superH}) takes
the form
\begin{equation}
{\cal H} = - \frac{p_a^2}{24a}-6ka  +a^3
\rho_f+A^{\frac{1}{1+\alpha}}
a^3+\frac{A^{\frac{\alpha}{1+\alpha}}}{1+\alpha}a^{-3\alpha}S
p_{\Phi}^{1+\alpha},
\end{equation}
The following additional canonical transformations
\begin{eqnarray}
\hspace{-5mm}T
=-(1+\alpha)A^{-\frac{\alpha}{1+\alpha}}p_{\Phi}^{-(1+\alpha)}p_S,
\,\, p_T =\frac{A^{\frac{\alpha}{1+\alpha}}}{1+\alpha}S
p_{\Phi}^{1+\alpha},
\end{eqnarray}
simplify the super-Hamiltonian to
\begin{equation}
{\cal H} = - \frac{p_a^2}{24a} -6ka +Ba^{-3w}+A^{\frac{1}{1+\alpha}}
a^3+ a^{-3\alpha} p_T.\label{EqHamiltonian-b}
\end{equation}
The classical dynamics is governed by the Hamilton equations,
derived from Eq. (\ref{EqHamiltonian}) and Poisson brackets as
\begin{equation}
\left\{
\begin{array}{llllll}
\dot{a} =&\{a,N{\cal H}\}=-\frac{\displaystyle Np_{a}}{\displaystyle 12a} ,\\
 & \\
\dot{p_{a}} =&\{p_{a},N{\cal H}\}=- \frac{N}{24a^2}p_a^2+6Nk+3wNBa^{-3w-1}\\
&\\&
-3NA^{\frac{1}{1+\alpha}}a^2+3\alpha N \,a^{-3\alpha-1}p_T, \\
& \\
\dot{T} =&\{T,N{\cal H}\}=Na^{-3\alpha}\, ,\\
 & \\
\dot{p_{T}} =&\{p_{T},N{\cal H}\}=0\, .\\
& \\
\end{array}
\right. \label{4-b}
\end{equation}
We also have the constraint equation ${\cal H} = 0$. Choosing the
gauge $N=a^{3\alpha}$, we have the following solutions for the
system
\begin{eqnarray}\label{class1b}
T&=&t,\\
p_T&=&\textrm{const.},\\
\ddot{a}&=&(3\alpha-\frac{1}{2})\frac{\dot
a^2}{a}-\frac{k}{2}a^{6\alpha-1}-\frac{1}{4}wBa^{6\alpha-3w-2}
+\frac{1}{4}A^{\frac{1}{1+\alpha}}a^{6\alpha+1}-\frac{1}{4}\alpha
p_T a^{3\alpha-2},\\
0&=&-6a^{-6\alpha+1}\dot a^2-6k a +Ba^{-3w} +A^{\frac{1}{1+\alpha}}
a^3+a^{-3\alpha}\,p_T.
\end{eqnarray}
It is important to note that these equations predict an accelerating
Universe for late times. For large values of the scale factor we can
simplify the above equations and find the acceleration parameter
\begin{equation}
q=\frac{a\ddot{a}}{\dot{a}^2}=3\alpha-1,
\end{equation}
which is positive for $\alpha>1/3$. Now, imposing the standard
quantization conditions on the canonical momenta and demanding that
the super-Hamiltonian operator annihilates the wave function, we are
led to SWD equation in minisuperspace ($\hbar =1$)
\begin{equation}
\label{sle-b} \frac{\partial^2\Psi}{\partial a^2} -
(144ka^2-24Ba^{1-3w}-24A^{\frac{1}{1+\alpha}} a^4)\Psi -
i24a^{1-3\alpha}\frac{\partial\Psi}{\partial t} = 0 .
\end{equation}
Here, according to (\ref{class1b}), $T=t$ corresponds to the time
coordinate.  Demanding that the Hamiltonian operator ${\hat H}$ to
be self-adjoint, the inner product of any two wave functions $\Phi$
and $\Psi$ must take the form \cite{nivaldo, 15}
\begin{equation}
(\Phi,\Psi) = \int_0^\infty a^{1-3\alpha}\,\Phi^*\Psi da.
\end{equation}
The SWD equation (\ref{sle-b}) can  be solved by separation of
variables as follows
\begin{equation}
\psi(a,t) = e^{iEt}\psi(a), \label{11-b}
\end{equation}
where the $a$ dependent part of the wave function $\psi(a)$
satisfies
\begin{eqnarray}
\label{sle2-b} -\psi''(a) +\left(144 ka^2-24Ba^{1-3w}
-24A^{\frac{1}{1+\alpha}} a^4\right)\psi(a)
=24Ea^{1-3\alpha}\,\psi(a),
\end{eqnarray}
and the prime means derivative with respect to $a$. Note that
effective Chaplygin gas term ($24A^{\frac{1}{1+\alpha}}$) plays the
role of a positive cosmological constant. In particular, when
$\alpha=1/3$ and $w=1/3$, this equation reduces to the FRW model
with positive cosmological constant and radiation which has been
studied in Ref.~\cite{monerat}.

\section{Results}\label{sec3}
In this Section we first study the issue of singularity avoidance in
quantum cosmology in the early Universe and then present some
analytical solutions in both early and late Universes.

\begin{figure}
\centering
  % Requires \usepackage{graphicx}
  \includegraphics[width=8cm]{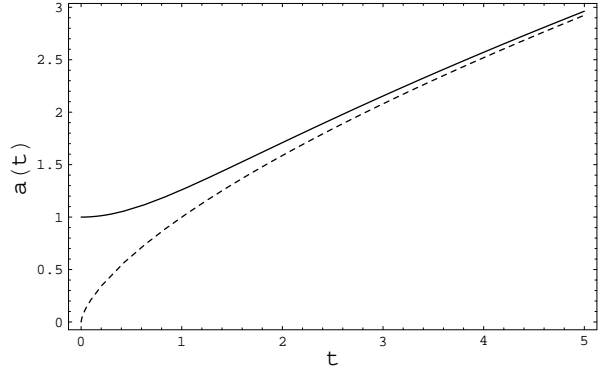}\\
  \caption{The time behavior of the expected value
for the scale factor $\langle a\rangle(t)$ (solid line) and the
classical scale factor $a(t)$ (dashed line) for dust dominated
Universe ($w=0$) and flat space time ($k=0$).}\label{fig1}
\end{figure}

For $k = 0$ the time-independent Wheeler-DeWitt equation
(\ref{sle2}), in the dust dominated Universe ($w=0$), reduces to
\begin{equation}
\label{eq-dust} \psi'' + 24(E+B)a\psi = 0.
\end{equation}
The above equation has the following general time-dependent
solutions under the form of Bessel functions
\begin{equation} \label{bessel} \Psi_E' =
e^{iEt}\sqrt{a}\biggr[c_1J_{\frac{1}{3}}\biggr(\frac{\sqrt{96E'}}{3}a^{\frac{3}{2}}\biggl)
+
c_2Y_{\frac{1}{3}}\biggr(\frac{\sqrt{96E'}}{3}a^{\frac{3}{2}}\biggl)\biggl],
\end{equation}
where $E'=E+B$. Now, the wave packets can be constructed by
superimposing these solutions to obtain physically allowed wave
functions. The general structure of these wave packets are
\begin{equation}
\Psi(a,t) = \int_0^\infty A(E')\Psi_E'(a,t)dE'  .
\end{equation}
We choose $c_2 = 0$ for satisfying the first boundary condition
(\ref{boundary}). Defining $r =\frac{\sqrt{96E'}}{3}$, simple
analytical expressions for the wave packet can be found by choosing
$A(E')$ to be a quasi-gaussian function
\begin{equation}
\Psi(a,t) = \sqrt{a}e^{-iBt}\int_0^\infty r^{\nu + 1}e^{-\gamma r^2
+ i\frac{3}{32}r^{2} t}J_\nu(ra^\frac{3}{2})dr,
\end{equation}
where $\nu = \frac{1}{3}$ and $\gamma$ is an arbitrary positive
constant. The above integral is known \cite{gradshteyn}, and the
wave packet takes the form
\begin{equation}
\label{wp} \Psi(a,t) =
a\frac{e^{-\frac{a^{3}}{4Z}-iBt}}{(-2Z)^{\frac{4}{3}}},
\end{equation}
where $Z=\gamma-i\frac{3}{32}t$. Now, we can verify what these
quantum models predict for the behavior of the scale factor of the
Universe. By adopting the many-worlds interpretation
\cite{tipler,everett}, and with regards to the inner product
relation (\ref{inner}), the expectation value of the scale factor
\begin{equation}
<a>(t) = \frac{\int_0^\infty a\Psi(a,t)^*a\Psi(a,t)da}
{\int_0^\infty a\Psi(a,t)^*\Psi(a,t)da},
\end{equation}
is easily computed, leading to
\begin{equation} <a>(t) \propto
\biggr[\frac{9}{(32)^2\gamma^2}t^2 + 1\biggl]^\frac{1}{3} .
\end{equation}
These solutions represent a no singular Universe which goes
asymptotically over to the corresponding flat classical model for
dust ($w=0$) dominated epoch
(\ref{class1}-\ref{class4})(Fig.~\ref{fig1})
\begin{equation}
a(t) \propto t^{2/3}.
\end{equation}

In the case $k=1$ and $w=0$ the time-independent Wheeler-DeWitt
equation (\ref{sle2}) reduces to
\begin{equation}
-{\psi}^{\prime \prime}(a) + \left(- 24E'a +
144a^{2}\right){\psi}(a)=0.
\end{equation}
Defining new variable $x=12a - E'$ we find
\begin{equation}\label{k1}
-\frac{d^{2}\psi}{dx^{2}}+\left[-
\frac{E'^{2}}{144}+\frac{x^{2}}{144} \right]\psi(a) =0.
\end{equation}
Equation (\ref{k1}) is similar to the time-independent Schr\"odinger
equation for a simple harmonic oscillator with unit mass and energy
$\lambda$
\begin{equation}
-\frac{d^{2}\psi}{dx^{2}}+\left[- 2\lambda+w^{2}x^{2}\right]\psi(x) =0,
\end{equation}
where $2\lambda = E'^{2}/144$ and $w=1/12$. Therefore, the allowed
values of $\lambda$ are $w(n+1/2)$ and the possible values of $E'$
are
\begin{equation}
E'_{n}=\sqrt{12(2n+1)}\,\, , \mbox{\hspace{0.8cm}} n=0,1,2,...\quad.
\end{equation}
therefore, the  stationary solutions are
\begin{equation}
{\Psi}_{n}(a,t)=e^{iE_{n}t}{\varphi}_{n}\left(12a - E'_{n}\right),
\label{k1-final}
\end{equation}
where
\begin{equation}
{\varphi}_{n}(x)=H_n\bigg(\frac{x}{\sqrt{12}}\bigg)e^{-x^2/24}\,\, ,
%\frac{(-1)^{n}}{\sqrt{2^{n}n!\sqrt{n}}}e^{x^{2}/2}\frac{d^{n}}{dx^{n}}\left( e^{-x^{2}}\right) , \mbox{\hspace{0.8cm}} n=0,1,2,...
\label{dust7}
\end{equation}
and $H_n$ are Hermite polynomials. The wave functions
(\ref{k1-final}) are similar to the stationary quantum wormholes as
defined in \cite{Hawking}. However, neither of the boundary
conditions (\ref{boundary}) can be satisfied by the these wave
functions.

In $k=-1$ and $w=0$ case, equation (\ref{sle2}) reduces to
\begin{equation}
{\psi}^{\prime \prime}(a) + \left(24E'a +
144a^{2}\right){\psi}(a)=0,
\end{equation}
where the solutions are
\begin{eqnarray}
\label{whittaker} \Psi (a,t)=e^{iEt}(12a+E')^{-1/2}\bigg\{
C_{1}M_{\frac{iE^2}{48},\frac{1}{4}}\left(\frac{i(12a+E')^2}{12}\right)+
C_{2}W_{\frac{iE^2}{48},\frac{1}{4}}\left(\frac{i(12a+E')^2}{12}\right)\bigg\},
\end{eqnarray}
where $M_{\kappa , \lambda}$ and $W_{\kappa , \lambda}$ are
Whittaker functions. The Whittaker functions do not automatically
vanish at $a=0$. Therefore, we need to take both $C_{1}\neq 0$ and
$C_{2}\neq 0$ to satisfy $\Psi(0,t)=0$.

For $w=-1/3$, the SWD equation (\ref{sle2}) can be written as
\begin{equation}
-\psi''(a) +24(6 k-B)a^2\psi(a) =24Ea\,\psi(a),
\end{equation}
which as before, has the solutions in the form of Simple Harmonic
Oscillator (\ref{k1-final}) with discrete spectrum or Whittaker
function (\ref{whittaker}) for positive or negative value of $(6
k-B)$, respectively.

%*********************************************************************************************

For $k=0$ and $w=1/3$ (radiation), the WD equation (\ref{sle2})
reduces to
\begin{equation}
\label{radiation-1} -\psi''(a) -24B\psi(a) =24Ea\psi(a),
\end{equation}
which can be rewritten as
\begin{equation}
\psi''(a) +24E\left(a+ \frac{\displaystyle B}{\displaystyle
E}\right)\psi(a) =0,
\end{equation}
by taking $x= a+\frac{\displaystyle B}{\displaystyle E}$ we have
\begin{equation}
\frac{d^2}{dx^2}\psi(x) +24 E x\psi(x) =0,
\end{equation}
which is the Airy's differential equation. We solve this equation
for $E>0$ and $E<0$, separately.

For $E>0$, this equation has two solutions as $\mbox{Ai}\left[-(24
E)^{1/3}x\right]$ and $\mbox{Bi}\left[-(24 E))^{1/3}x\right]$. First
one is exponentially decreasing function of $x$ and the second one
grows exponentially and is physically unacceptable. Therefore, the
solution is
\begin{equation}
\psi(a)= \mbox{Ai}\left[-(24 E))^{1/3}\left(a+\frac{\displaystyle
B}{\displaystyle E}\right)\right].
\end{equation}
We choose the first boundary condition (\ref{boundary}), which leads
to
\begin{equation}
\mbox{Ai}\left[{ -}{ \left(24E\right)^{1/3}}\frac{\displaystyle
B}{\displaystyle E}\right]=0.
\end{equation}
Airy's function $\mbox{Ai}(x)$ has infinitely many negative zeros
$z_n = -a_n$, where $a_n>0$, therefore, the energy levels quantize
and take the values
\begin{equation}
E_n = \left(\frac{{24}^{1/3}B}{a_n}\right)^{3/2}.
\end{equation}
The time-dependent eigenfunctions take the form
\begin{equation}
\Psi_n(a,t)=e^{iE_n
t}\mbox{Ai}\left[-(24E_n)^{1/3}\left(a+\frac{\displaystyle
B}{\displaystyle E_n}\right)\right].
\end{equation}

\begin{figure}
\centering
  % Requires \usepackage{graphicx}
  \includegraphics[width=8cm]{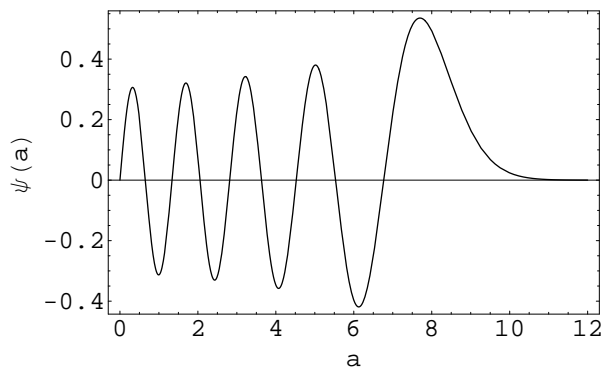}\\
  \caption{Plot of the wave function ($\psi(a)$) for $B=1$ and $n=8$, showing the oscillatory behavior for the small values of the scale factor
  and exponential damping for the large values of the scale factor.}\label{fig2}
\end{figure}

For $E<0$, this equation has also two solutions as
$\mbox{Ai}\left[(24 |E|)^{1/3}x\right]$ and $\mbox{Bi}\left[(24
|E|))^{1/3}x\right]$. Since the second one grows exponentially and
is physically unacceptable, the solution is
\begin{equation}
\psi(a)= \mbox{Ai}\left[(24 |E|))^{1/3}\left(a-\frac{\displaystyle
B}{\displaystyle |E|}\right)\right].
\end{equation}
We choose the first boundary condition (\ref{boundary}), which leads
to
\begin{equation}
\mbox{Ai}\left[-{ \left(24|E|\right)^{1/3}}\frac{\displaystyle
B}{\displaystyle |E|}\right]=0,
\end{equation}
therefore, the energy levels quantize and take the values
\begin{equation}
E_n = -\left(\frac{{24}^{1/3}B}{a_n}\right)^{3/2}.
\end{equation}
The time-dependent eigenfunctions take the form
\begin{equation}\label{radiation-final-1}
\Psi_n(a,t)=e^{iE_n
t}\mbox{Ai}\left[(24|E_n|)^{1/3}\left(a-\frac{\displaystyle
B}{\displaystyle |E_n|}\right)\right].
\end{equation}
It is important to note that Airy's function $ \mbox{Ai}(x)$ has an
oscillatory behavior for $x<0$ ($a<\frac{\displaystyle
B}{\displaystyle |E_n|}$) whiles for $x>0$ ($a>\frac{\displaystyle
B}{\displaystyle |E_n|}$) decreases monotonically and is an
exponentially damped function for large $x$ (Fig.~\ref{fig2}).
Therefore, the solutions (\ref{radiation-final-1}) show a classical
behavior for small $a$ and a quantum behavior for large $a$. This is
contrary to usually expected results for previous case. In fact
detecting quantum gravitational effects in large Universes is
noticeable which has been also observed in FRW, Stephani, and
Kaluza-Klein models \cite{lemos1999,pedramCQG2,Coliteste}.

In $k=1$ and $w=1/3$ (radiation) case, the WD equation (\ref{sle2})
reduces to
\begin{equation}
\label{cosmic-1} -\psi''(a) + (144 a^2-24B)\psi(a) =24Ea\psi(a).
\end{equation}
The above equation can be written as
\begin{equation}
-\psi''(a) +144
\left[\left(a-\frac{E}{12}\right)^2-\left(\frac{E}{12}\right)^2\right]\psi
=0,
\end{equation}
by taking $x=a- \frac{\displaystyle E}{\displaystyle 12}$ we have
\begin{equation}
-\frac{d^2}{dx^2}\psi(x) + 144x^2\psi(x) =(E^2+24B)\psi(x).
\end{equation}
This equation is identical to the time-independent Schr\"odinger
equation for a simple harmonic oscillator with unit mass and energy
$\lambda$
\begin{equation}
-\frac{d^{2}\psi(x)}{dx^{2}}+\omega^{2}x^{2}\psi(x) =2\lambda
\psi(x),
\end{equation}
where $2\lambda = (E^2+24B)$ and $\omega^2=144$. Therefore, the
allowed values of $\lambda$ are $\omega(n+1/2)$ and the possible
values of $E$ are
\begin{equation}
E_{n}=\sqrt{6(n+1/2)-24B}\,\, , \mbox{\hspace{0.8cm}}
n=0,1,2,...\quad ,
\end{equation}
therefore, the stationary solutions are
\begin{equation}
{\Psi}_{n}(a,t)=e^{iE_{n}t}{\varphi}_{n}\left(a- \frac{\displaystyle
E_n}{\displaystyle 12}\right),
\end{equation}
\begin{equation}
{\varphi}_{n}(x)=H_n\left((12)^{\frac{1}{2}}x\right)e^{-3\,\,x^2},
\end{equation}
where $H_n$ are Hermite polynomials.  However, neither of the
boundary conditions (\ref{boundary}) can be satisfied by these wave
functions.

%*********************************************************************************************

Now, we present some analytical solutions for the late time
Universe. For flat space time ($k=0$), dust epoch ($w=0$), and
standard Chaplygin gas ($\alpha=1$), equation (\ref{sle2-b}) reduces
to
\begin{equation}
-\psi''(a) +\left(-24Ba -24A^{\frac{1}{1+\alpha}} a^4\right)\psi(a)
=24Ea^{-2}\,\psi(a),
\end{equation}
where the solutions are
\begin{eqnarray}\nonumber
\Psi (a,t)&=&e^{iEt}e^{2 i \sqrt{\frac{2}{3}}
A^{\frac{1}{2(1+\alpha)}} a^3}
   a^{\frac{1}{2}-\frac{1}{2} \sqrt{1-96
   E}}\\ \nonumber &\times&\bigg\{C_1\,\,
U\left(\frac{1}{6} \left(-2 i
   \sqrt{6} BA^{\frac{-1}{2(1+\alpha)}}-\sqrt{1-96
   E}+3\right),1-\frac{1}{3} \sqrt{1-96
   E},-4 i \sqrt{\frac{2}{3}} A^{\frac{1}{2(1+\alpha)}}
   a^3\right)\\ &+&C_2\,\,
   L_{\frac{1}{6} (2 i \sqrt{6}
   BA^{\frac{-1}{2(1+\alpha)}}+\sqrt{1-96
   E}-3)}^{-\frac{1}{3} \sqrt{1-96
   E}}\left(-4 i \sqrt{\frac{2}{3}} A^{\frac{1}{2(1+\alpha)}}
   a^3\right)\bigg\}.
\end{eqnarray}
Here $U(a,b,c)$ is the confluent hypergeometric function and
$L_n^a(x)$ is the generalized Laguerre polynomial. We need to take
both $C_{1}\neq 0$ and $C_{2}\neq 0$ to satisfy $\Psi(0,t)=0$.

For flat space time ($k=0$), stiff matter ($w=1$), and standard
Chaplygin gas ($\alpha=1$), equation (\ref{sle2-b}) reduces to
\begin{equation}
-\psi''(a) +\left(-24Ba^{-2} -24A^{\frac{1}{1+\alpha}}
a^4\right)\psi(a) =24Ea^{-2}\,\psi(a),
\end{equation}
with the solutions as
\begin{eqnarray}\nonumber
\Psi (a,t)&=&e^{iEt}\bigg\{C_1\,\,\sqrt{a} \,J_{-\frac{1}{6}
\sqrt{-96( B+
   E)+1}}\left(2 \sqrt{\frac{2}{3}} A^{\frac{1}{2(1+\alpha)}}
   a^3\right)\\ &+&C_2\,\,\sqrt{a}\, J_{\frac{1}{6} \sqrt{-96( B+
   E)+1}}\left(2 \sqrt{\frac{2}{3}} A^{\frac{1}{2(1+\alpha)}}
   a^3\right)\bigg\}.
\end{eqnarray}
Here again, we have $C_{1}\neq 0$ and $C_{2}\neq 0$ in order to
satisfy the first boundary condition (\ref{boundary}).

\section{Conclusions}\label{sec4}
In this work we have investigated minisuperspace FRW quantum
cosmological models with Chaplygin gas and perfect fluid as the
matter content in early and late times. The use of Schutz's
formalism for the Chaplygin gas allowed us to obtain SWD equations
with the perfect fluid's effective potential. We have obtained
eigenfunctions and therefore acceptable wave packets were
constructed by appropriate linear combination of these
eigenfunctions. The time evolution of the expectation value of the
scale factor has been determined in the spirit of the many-worlds
interpretation of quantum cosmology. We have showed that contrary to
the classical case, the expectation value of the scale factor avoids
singularity at the quantum level. Moreover, this model predicts an
accelerated Universe for late times.


\begin{thebibliography}{100}
\bibitem{Riess:1998cb} A.~G.~Riess {\it et al.}  [Supernova Search Team Collaboration], Astron.~J.~{\bf 116}, 1009 (1998) [arXiv:astro-ph/9805201]; S.~Perlmutter {\it et al.}  [Supernova Cosmology Project Collaboration], Astrophys.~J.~ {\bf 517}, 565 (1999) [arXiv:astro-ph/9812133]; J.~L.~Tonry {\it et al.}, Astrophys.~J.~{\bf 594}, 1 (2003) [arXiv:astro-ph/0305008].
\bibitem{Spergel:2003cb} D.~N.~Spergel {\it et al.}, Astrophys.~J.~Suppl.~ {\bf 148}, 175 (2003) [arXiv:astro-ph/0302209]; C.~L.~Bennett {\it et al.}, Astrophys.~J.~Suppl.~ {\bf 148}, 1 (2003) [arXiv:astro-ph/0302207].
\bibitem{2a} M.~Tegmark {\it et al.}  [SDSS Collaboration], [arXiv:astro-ph/0310723].
\bibitem{3a} V.~Sahni, Class.~Quant.~Grav.~ {\bf 19}, 3435 (2002) [arXiv:astro-ph/0202076].
\bibitem{3b} S.~Weinberg, Rev.~Mod.~Phys.~{\bf 61}, 1 (1989).
\bibitem{3c} P.~J.~E.~Peebles and B.~Ratra, Rev.~Mod.~Phys.~ {\bf 75}, 559 (2003) [arXiv:astro-ph/0207347].
\bibitem{Wetterich:fm} C.~Wetterich, Nucl.~Phys.~B {\bf 302}, 668 (1988); B.~Ratra and P.~J.~E.~Peebles, Phys.~Rev.~D {\bf 37}, 3406 (1988); R.~R.~Caldwell, R.~Dave and P.~J.~Steinhardt, Phys.~Rev.~Lett.~{\bf 80}, 1582 (1998) [arXiv:astro-ph/9708069]; P.~F.~Gonz\'{a}lez-D\'{i}az, Phys.~Rev.~D {\bf 62}, 023513 (2000) [arXiv:astro-ph/0004125]; Y.~Fujii, Phys.~Rev.~D {\bf62}, 064004 (2000) [arXiv:gr-qc/9908021].
\bibitem{Brax:1999yv} P.~Brax and J.~Martin, Phys.~Rev.~D {\bf61}, 103502 (2000) [arXiv:astro-ph/9912046].
\bibitem{Kamenshchik} A.~Y.~Kamenshchik, U.~Moschella and V.~Pasquier, Phys.~Lett.~B {\bf 511}, 265 (2001) [arXiv:gr-qc/0103004].
\bibitem{A} M.~C.~Bento, O.~Bertolami and A.~A.~Sen, Phys.~Rev.~D {\bf66}, 043507 (2002)  [arXiv:gr-qc/0202064].
\bibitem{Bento} M.~C.~Bento, O.~Bertolami and A.~A.~Sen, Phys.~Rev.~D {\bf67},  063003 (2003)  [arXiv:astro-ph/0210468].
\bibitem{3rp} M.~C.~Bento, O.~Bertolami and A.~A.~Sen, Phys.~Lett.~B {\bf 575}, 172  (2003) [arXiv:astro-ph/0303538]; L.~Amendola, F.~Finelli, C.~Burigana and D.~Carturan, JCAP {\bf 0307}, 005 (2003) [arXiv:astro-ph/0304325].
\bibitem{2rp} R.~Bean and O.~Dore, Phys.~Rev.~D {\bf 68}, 023515 (2003) [arXiv:astro-ph/0301308]; A.~Dev, D.~Jain and J.S.~Alcaniz, astro-ph/0311056; M.~Biesiada, W.~Godlowski and M.~Szydlowski, astro-ph/0403305.
\bibitem{Fabris} J.~C.~Fabris, S.~V.~Goncalves and P.~E.~de Souza, Gen.~Rel.~Grav.~ {\bf 34}, 53 (2002) [arXiv:gr-qc/0103083]; Gen.~Rel.~Grav.~ {\bf 34}, 2111 (2002) [arXiv:astro-ph/0203441]; T.~Multamaki, M.~Manera and E.~Gaztanaga, Phys.~Rev.~D {\bf 69}, 023004 (2004) [arXiv:astro-ph/0307533].
\bibitem{C} V.~Gorini, A.~Kamenshchik and U.~Moschella, Phys.~Rev.~D {\bf67}, 063509 (2003)  [arXiv:astro-ph/0209395]; R.~Colistete, J.~C.~Fabris, S.~V.~Goncalves and P.~E.~de Souza, [arXiv:gr-qc/0210079], H.~Sandvik, M.~Tegmark, M.~Zaldarriaga and I.~Waga, [arXiv:astro-ph/0212114]; L.~M.~Beca, P.~P.~Avelino, J.~P.~de Carvalho and C.~J.~Martins, Phys.~Rev.~D {\bf67}, 101301 (2003)  [arXiv:astro-ph/0303564].
\bibitem{Ogawa} N.~Ogawa, Phys.~Rev.~D {\bf62}, 085023 (2000)  [arXiv:hep-th/0003288].
\bibitem{NewBD} M.~Bordemann and J.~Hoppe, Phys.~Lett.~B {\bf 317}, 315 (1993) [arXiv:hep-th/9307036].
\bibitem{18a} M.~Hassaine and P.~A.~Horvathy, Lett.~Math.~Phys.~ {\bf 57}, 33 (2001) [arXiv:hep-th/0101044].
\bibitem{20a} G.~W.~Gibbons, Grav.~Cosmol.~ {\bf 8}, 2 (2002) [arXiv:hep-th/0104015].
\bibitem{21a} M.~Hassaine, Phys.~Lett.~A {\bf 290}, 157 (2001) [arXiv:hep-th/0106252].
\bibitem{23a} G.~M.~Kremer, Gen.~Rel.~Grav.~{\bf 35}, 1459 (2003) [arXiv:gr-qc/0303103].
\bibitem{27a} H.~B.~Benaoum, [arXiv:hep-th/0205140].
\bibitem{Rev1} V.~Gorini, A.~Kamenshchik, U.~Moschella and V.~Pasquier, [arXiv:gr-qc/0403062].
\bibitem{Rev2} O.~Bertolami, [arXiv:astro-ph/0403310].
\bibitem{DySy} M.~Szydlowski and W.~Czja, Phys.~Rev.~D {\bf 69}, 023506 (2004) [arXiv:astro-ph/0306579].
\bibitem{Jackiw} R.~Jackiw, [arXiv:physics/0010042].
\bibitem{setare1} M.~R.~Setare, Phys.Lett.~B 644, 99 (2007).
\bibitem{setare2} M.~R.~Setare, Phys.~Lett.~B 648, 329 (2007).
\bibitem{setare3} M.~Roos, [arXiv:0704.0882].
\bibitem{Buahmadi} M.~Bouhmadi-L\'{o}pez, P.~V.~Moniz, Phys.~Rev.~D \textbf{71}, 063521 (2005).
\bibitem{chap} P.~Pedram, S.~Jalalzadeh and S.~S.~Gousheh, Int.~J.~Theor.~Phys.~DOI: 10.1007/s10773-007-9436-9  [arXiv:0705.3587].
\bibitem{11} B.~F.~Schutz, Phys.~Rev.~D {\bf  2}, 2762 (1970).
\bibitem{12} B.~F.~Schutz, Phys.~Rev.~D {\bf  4}, 3559 (1971).
\bibitem{PLB} P.~Pedram, S.~Jalalzadeh and S.~S.~Gousheh, Phys.~Lett.~B.~In press, doi:10.1016/j.physletb.2007.08.077, [arXiv:0708.4143].
\bibitem{pedramCQG2} P.~Pedram, S.~Jalalzadeh and S.~S.~Gousheh, Class.~Quantum Grav.~In press, [arXiv:0709.1620].
\bibitem{FRW} F.~G.~Alvarenga, J.~C.~Fabris, N.~A.~Lemos, and G.~A.~Monerat,  Gen.~Rel.~Grav.~\textbf{34} 651 (2002).
\bibitem{monerat2} G.~A.~Monerat, G.~Oliveira-Neto, E.~V.~Corr\^{e}a Silva, L.~G.~Ferreira Filho, P.~Romildo, Jr., J.~C.~Fabris, R.~Fracalossi, S.~V.~B.~Gon\c{c}alves, and F.~G.~Alvarenga, Phys.~Rev.~D \textbf{76}, 024017 (2007)
\bibitem{ref1} O.~Bertolami and J.~Mourao, Class Quantum Grav.~\textbf{8}, 1271 (1991).
\bibitem{ref2} O.~Bertolami and V.~Duvvuri, Phys.~Lett.~B \textbf{640}, 121 (2006).
\bibitem{ref3} T.~Barreiro, A.A.~Sen, Phys.~Rev.~D \textbf{70}, 124013 (2004).
\bibitem{ref4} M.~Heydari-Fard and H.~R.~Sepangi, to appear in Phys.~Rev.~D, [arXiv: 0710.2666].
\bibitem{ref5} S.~Jalalzadeh and H.~R.~Sepangi, Class.~Quant Grav.~\textbf{22}, 2035 (2005).
\bibitem{ref7} G.~F.~R.~Ellis and S.~W.~Hawking, Large Scale Structure of Space Time, (Cambridge University Press, 1973); R.~Mansouri and F.~Nasseri, Phys.~Rev.~D \textbf{60}, 123512 (1999).
\bibitem{7} R.~Arnowitt, S.~Deser and C.~W.~Misner, {\it Gravitation: An Introduction to Current Research}, edited by L.~Witten, Wiley, New York (1962).
\bibitem{14} V.~G.~Lapchinskii and V.~A.~Rubakov, Theor.~Math.~Phys.~{\bf 33}, 1076 (1977).
\bibitem{nivaldo} N.~A.~Lemos, J.~Math.~Phys.~{\bf 37}, 1449 (1996).
\bibitem{15} F.~G.~Alvarenga and N.~A.~Lemos, Gen.~Rel.~Grav.~{\bf 30}, 681 (1998).
\bibitem{monerat} J.~Acacio de Barros, E.~V.~Corr\^{e}a Silva, G.~A.~Monerat, G.~Oliveira-Neto, L.~G.~Ferreira Filho, and P.~Romildo, Phys.~Rev.~D \textbf{75}, 104004 (2007).
\bibitem{gradshteyn} I.~S.~Gradshteyn and I.~M.~Ryzhik, {\it Table of Integrals, Series and Products} (Academic, New York, 1980), formula 6.631-4.
\bibitem{tipler} F.~J.~Tipler, Phys.~Rep.~{\bf 137}, 231 (1986).
\bibitem{everett} H.~Everett, III, Rev.~Mod.~Phys.~{\bf 29}, 454 (1957).
\bibitem{Hawking} S.~W.~Hawking and D.~B.~Page, Phys.~Rev.~D \textbf{42}, 2655 (1990).
\bibitem{lemos1999} N.~A.~Lemos, F.~G.~Alvarenga, Gen.~Rel.~Grav.~\textbf{31}, 1743 (1999), [arXiv:gr-qc/9906061].
\bibitem{Coliteste} R.~Coliteste, Jr., J.~C.~Fabris and N.~Pinto-Neto,  Phys.~Rev.~D {\bf57}, 4707 (1998).
\end{thebibliography}
\end{document}